# Observation of two-component exciton condensates in an excitonic insulator

Ruishi Qi[1,2,†], Qize Li[1,3,†], Jiahui Nie[1,2,3], Ruichen Xia[1], Haleem Kim[1,2], Hyungbin Lim[1,2], Jingxu Xie[1,2,3], Takashi Taniguchi[4], Kenji Watanabe[5], Michael F. Crommie[1,2], Allan H. MacDonald[6], Feng Wang[1,2,7,*]

[1] Department of Physics, University of California, Berkeley, CA 94720, USA.
[2] Materials Sciences Division, Lawrence Berkeley National Laboratory, Berkeley, CA 94720, USA.
[3] Graduate Group in Applied Science and Technology, University of California, Berkeley, CA 94720, USA.
[4] Research Center for Materials Nanoarchitectonics, National Institute for Materials Science, 1-1 Namiki, Tsukuba 305-0044, Japan.
[5] Research Center for Electronic and Optical Materials, National Institute for Materials Science, 1-1 Namiki, Tsukuba 305-0044, Japan.
[6] Department of Physics, University of Texas at Austin, Austin, TX 78712, USA.
[7] Kavli Energy NanoScience Institute, University of California Berkeley and Lawrence Berkeley National Laboratory, Berkeley, CA 94720, USA.
[†] These authors contributed equally to this work.
[*] To whom correspondence should be addressed: fengwang76@berkeley.edu.

**Macroscopic quantum coherence emerges when bosons condense into a Bose-Einstein condensate (BEC). First observed as a single-component superfluid in helium[1,2], BECs later emerged in ultracold atomic gases at nanokelvin temperatures as weakly interacting quantum fluids[3,4], which can also host multicomponent spinor condensates with rich internal degrees of freedom[5]. Excitons provide a promising solid-state platform for BECs that can combine strong interactions, electrical tunability, high transition temperatures, and multicomponent order. Yet, conclusive evidence for condensation has remained elusive. Here, we report evidence of two-component exciton BECs in MoSe$_2$/hBN/WSe$_2$ electron-hole bilayers[6–9] by directly probing the spin susceptibility of constituent electrons and holes. This heterostructure hosts equilibrium exciton fluids with four spin-valley flavors. Using magneto-optical spectroscopy in a dilution refrigerator, we reveal three exciton condensate phases with distinct flavor polarizations. At zero magnetic field, the many-body ground state is a coherent superposition of two simultaneously condensed intravalley exciton flavors. Under a magnetic field, the intravalley exciton condensate first switches to a two-component intervalley exciton condensate via a first-order quantum phase transition at a weak critical field, and then turns into a fully-polarized single-component condensate at high fields. The two-component condensates persist up to ~1.8 K. Our results establish van der Waals electron-hole bilayers as a versatile platform for exploring strongly interacting, multicomponent exciton BECs.**

Exemplified by superfluid helium[1,2] and atomic Bose-Einstein condensates (BECs)[3,4], macroscopic quantum coherence arises when bosons condense into a single quantum state. Over 60 years ago, Keldysh theorized that excitons can provide an attractive platform to realize high-temperature BECs[10]. Exciton BECs combine several appealing characteristics[11–15]: electron-hole (e-h) interactions can be very strong, the exciton density is easily tunable, and, unlike most bosonic fluids, excitons can carry multiple internal degrees of freedom inherited from the spin and valley of their constituents. As a result, excitons could form multicomponent BECs with rich internal symmetries, reminiscent of spinor condensates in ultracold atoms[5] and the multicomponent superfluidity of $^3$He (refs.[16,17]).

Despite the promise, their experimental realization has remained elusive after decades of sustained endeavor. Optically generated excitons are too short-lived for equilibrium condensation[18,19]. Bulk excitonic insulator (EI) candidates usually exhibit competing lattice instabilities[20]. Quantum Hall bilayers based on GaAs and graphene have demonstrated interlayer phase coherence that can be viewed as magneto-exciton condensation in a particle-hole-transformed picture[21–26], but it exists only within quantum Hall states under high magnetic fields, and the spin degree of freedom is frozen.

E-h bilayers built from transition metal dichalcogenides (TMDs) have recently emerged as a powerful system for realizing equilibrium interlayer excitons at zero magnetic field[6–9,12,13,15,27]. Confining a two-dimensional electron gas (2DEG) and a 2D hole gas (2DHG) in two closely spaced but electrically isolated layers, such heterostructures host strongly correlated exciton fluids that can be tuned electrostatically. Optical, capacitance, and Coulomb drag measurements have established the existence of tightly bound indirect excitons in this equilibrium ground state[6–9,28,29]. However, these studies do not distinguish a coherent condensate from a classical exciton gas lacking phase coherence. While counterflow transport is regarded as the hallmark of exciton superfluidity, it is experimentally inaccessible in current TMD bilayers due to challenges in making electrical contacts to the excitons. Direct evidence of the long-sought exciton BEC in e-h bilayers has remained missing.

Here we report experimental evidence of two-component exciton BECs in $MoSe_2/hBN/WSe_2$ e-h bilayers. This heterostructure supports tunable exciton fluids with four internal spin-valley flavors, including two intravalley excitons with the electron and hole in the same valley and two intervalley excitons in the opposite valleys. (The electron and hole spin directions are locked with their valleys in TMDs.) Using magneto-optical spectroscopy in a dilution refrigerator, we probe the spin-valley susceptibility of the constituent electrons and holes of the exciton fluid, which exhibit distinct susceptibility behaviors in different BEC phases. We show that the condensate ground state in 60°-aligned $MoSe_2/hBN/WSe_2$ heterostructures is a two-component BEC of intravalley excitons, which exhibit a large flavor stiffness with small electron and hole spin susceptibilities. A weak magnetic field can drive the system to a two-component BEC of intervalley excitons through a first-order quantum phase transition. At a high magnetic field, a single-component BEC of intervalley excitons becomes the ground state. The Berezinskii-Kosterlitz-Thouless (BKT) transition temperature increases with the exciton density, reaching a maximum of ~1.8 K near the exciton Mott transition.

## Quantum-degenerate exciton fluid

The dual-gated e-h bilayer device is schematically shown in Fig. 1a. Similar structures have been described in detail in refs.[6–9,28–31]. It comprises a monolayer $MoSe_2$ electron layer and a monolayer $WSe_2$ hole layer, separated by a thin hBN tunneling barrier. The interlayer distance $d \approx 2$ nm provides a strong Coulomb attraction while exponentially suppressing interlayer tunneling. They are sandwiched by graphite top gate (TG) and bottom gate (BG) with hBN dielectrics, enabling independent electrostatic control of both layers through separate contacts. Applying an interlayer bias voltage $V_B \equiv V_h - V_e$ can overcome the type-II bandgap, simultaneously populating electrons in MoSe$_2$ and holes in WSe$_2$. Strong Coulomb attraction spontaneously binds them into stable dipolar excitons with density controllable by $V_B$. The symmetric gate voltage $V_G \equiv (V_{TG} + V_{BG})/2$ sets the Fermi level and controls e-h density imbalance.

Fig. 1b shows an optical image of device D1 with a 5-layer hBN spacer. While moiré effects are not expected, we observe a surprising twist-angle dependence in the exciton BEC phases. We will focus on device D1 with near-60° angle alignment in the main text. Data from over 10 devices are summarized in Extended Data Figs. 1-3. Unless noted otherwise, all measurements are performed at lattice temperature $T = 0.01$ K.

The strong intralayer exciton and trion resonances of monolayer TMDs enable direct optical readout of the carrier density and spin polarization[32,33]. Exploiting the sensitive doping dependence of the optical resonances, we can determine the electron and hole densities separately from the reflectance spectra[7]. Fig. 1c presents the charge doping phase diagram versus $V_G$ and $V_B$, where the red and green channels encode the measured electron ($n_e$) and hole ($n_h$) densities, respectively. For $V_B < 0.72$ V, the bilayer is a trivial band insulator (BI, black region). Adjusting $V_G$ moves the Fermi level into the valence or conduction band, forming a 2DHG in WSe$_2$ (green region) or a 2DEG in MoSe$_2$ (red region). Once $V_B$ exceeds the band offset, both layers are populated, forming a stable e-h fluid (yellow). The e-h bilayer is known to host an EI state at charge neutrality[6–9], which is charge incompressible as shown in Fig. 1d. Coulomb drag measurements further confirm its excitonic nature[8,9,28,29]. The charge gap, given by the width of the incompressible region, is strongly density-dependent. It starts from ~30 meV at low density and continuously decreases to zero at a critical density $n_{Mott} \approx 0.75 \times 10^{12}$ cm$^{-2}$ (Extended Data Fig. 4). Increasing $V_B$ beyond this point destroys the EI through an interaction-driven exciton Mott transition into a metallic electron-hole plasma (EHP)[6,7,9,34,35].

In TMD monolayers, strong spin-orbit coupling locks the spin and valley degrees of freedom. As illustrated in Fig. 1a (inset), the K-valley conduction and valence bands host spin-up electrons (note that hole spin is opposite to that of a valence electron), whereas the K' valley hosts spin-down electrons. Depending on the valley polarization of the constituent electron and hole, four interlayer exciton flavors are therefore possible. The spin-valley polarization of both electrons and holes can be measured using magnetic circular dichroism (MCD) enabled by valley-dependent optical selection rules[36–38]. Fig. 1e and 1f show the reflection contrast spectrum $\Delta R/R$ and MCD spectrum, respectively, at an exciton doping of $0.3 \times 10^{12}$ cm$^{-2}$. Each monolayer exhibits two characteristic intralayer resonances: the repulsive polaron peaks X$_0$ (intralayer exciton) and attractive polaron peaks X$^+$/X$^-$ (intralayer trion). Because the MoSe$_2$ and WSe$_2$ resonances lie at distinct wavelengths, we can obtain electron and hole polarizations separately. We denote the MCD amplitudes at the MoSe$_2$ X$^-$ and the WSe$_2$ X$^+$ resonances as eMCD (red arrow) and hMCD (green arrow), respectively, representing the spin-valley polarizations of electrons and holes[36,37]. This layer-resolved access is a unique advantage of interlayer excitons and is crucial for probing the internal degrees of freedom of exciton BECs.

Fig. 1g shows the eMCD signal across the doping phase diagram at 60 mT. A finite and relatively smooth eMCD appears whenever electrons are present, except for a pronounced enhancement inside the EI region and a suppression on its right side. The latter feature originates from spin-singlet interlayer trions discussed elsewhere[30,31], and will not be our focus here. Within the EI, the eMCD enhancement corresponds to a two- to threefold increase in electron spin susceptibility relative to a single-layer 2DEG. A similar enhancement of hole spin susceptibility is observed in the hMCD map in Fig. 1h.

The spin susceptibility in the EI phase provides key information on new spin-valley physics in the strongly interacting exciton fluid. Simple considerations predict very different susceptibilities for fermions and bosons due to their quantum statistics. In a fermionic 2DEG, spin polarization arises from the competition between the Zeeman energy $g\mu_B B$ ($g$, gyromagnetic ratio; $\mu_B$, Bohr magneton), which favors alignment, and the Fermi energy $E_F$, which opposes it. At $T = 0.01$ K, the thermal energy $k_B T$ (~μeV) is negligible compared with the meV-scale $E_F$ at typical carrier densities ($10^{11} - 10^{12} \text{cm}^{-2}$). The spin polarization, therefore, scales as $g\mu_B B/E_F$ – the standard Pauli paramagnetism. In the EI, however, all electrons and holes pair into bosonic excitons, removing $E_F$ from the competition. For a non-interacting exciton gas in the classical regime, the polarization should scale as $g\mu_B B/k_B T$, suggesting orders-of-magnitude enhancement at mK temperatures. For a non-interacting exciton gas in the BEC regime, the excitons are expected to form a fully polarized single-component BEC and become ferromagnetic[13]. Experimentally, we observe only a several-fold increase, which differs from either scenario. It suggests that strong exciton interactions, including interlayer e-h interactions and intralayer e-e and h-h interactions, generate a strongly correlated exciton fluid with non-trivial spin physics, as we describe below.

## Three exciton fluid phases with distinct spin-valley susceptibilities

Systematic measurements of the spin-valley polarization as a function of the exciton density and magnetic field $B$ reveal more exotic behaviors. We first compare the density- and field-dependent eMCD signals for electron-doping and exciton-doping scenarios. Fig. 2a shows the eMCD in the 2DEG phase with no hole present. The eMCD increases linearly with field, as expected for a Fermi liquid. Fig. 2b shows the eMCD at net charge neutrality (equal e-h densities), plotted as a function of field $B$ and exciton density $n_x$ ($= n_e = n_h$). While the eMCD signal at large $B$ grows rapidly, surprisingly, there is a white region near zero field with almost no eMCD. This feature is more evident in the spin susceptibility (i.e. $B$-derivative of the eMCD signal) map in Fig. 2c, where a nearly-unpolarizable gap spans several tens of millitesla. The eMCD slope inside the gap is not only strongly decreased – it actually changes sign and becomes slightly negative. We label this gap region as phase II$_A$, and the outside region as phase II$_B$. The phase II$_A$ persists to higher magnetic fields with increasing $n_x$ up to the exciton Mott transition at $0.75 \times 10^{12} \text{cm}^{-2}$, beyond which phase II$_A$ disappears. At the phase II$_A$/II$_B$ boundary, the spin susceptibility exhibits a peak, reminiscent of a first-order phase transition with a finite polarization jump. On a larger field range, the eMCD keeps growing until its saturation at tesla-scale fields, entering the fully polarized regime with zero susceptibility (labeled as phase I).

The hole susceptibility (i.e. $B$-derivative of hMCD) in the exciton fluid in Fig. 2d exhibits behavior strongly correlated with that of the electrons. It is overall negative across the entire field range (because the hole spin is opposite to that of a valence electron). The susceptibility magnitude is strongly reduced at small magnetic field (II$_A$ phase), but there is no sign-reversal. The susceptibility is enhanced in the II$_B$ phase until it enters hMCD saturation in phase I. Notably, phase boundaries between II$_A$, II$_B$, and I are identical in the electron (Fig. 2c) and hole (Fig. 2d) maps. Extended Data Fig. 5 provides a more complete dataset for the hMCD signals.

Fig. 2e-g show line cuts of the eMCD and hMCD at three different $n_x$ values. At low density ($n_x = 0.1 \times 10^{12} \text{cm}^{-2}$, Fig. 2e), the slope reduction associated with phase II$_A$ appears only weakly within a narrow field window. At high exciton density ($0.6 \times 10^{12} \text{cm}^{-2}$, Fig. 2f) the eMCD develops a slightly negative slope in phase II$_A$, corresponding to electron spins polarizing against the field. At the same time, the hMCD slope shows a pronounced but non-inverted suppression. For $n_x$ beyond the Mott transition, exciton melting produces a metallic EHP, and both MCD signals recover a simple linear behavior without anomalies (Fig. 2g, $n_x = 0.9 \times 10^{12} \text{cm}^{-2}$; here $n_x$ should be interpreted as e-h pair density rather than exciton density). Over a broader magnetic field range, eMCD and hMCD almost exactly mirror each other, as shown in Fig. 2h for exciton density $0.4 \times 10^{12} \text{cm}^{-2}$. Despite very different conduction and valence band g-factors ($g_c \approx 3, g_v \approx 6$)[39,40], electrons and holes reach full polarization at the same saturation field. (The saturated eMCD and hMCD signals correspond to fully polarized electrons and holes, respectively. They have different magnitudes owing to different optical signal strengths of the MoSe$_2$ and WSe$_2$ resonances.)

The correlated e-h responses imply a shared origin beyond single-particle physics. The singlet-like spin suppression in phase II$_A$ cannot arise from interlayer e-h hybridization within individual excitons because the hBN barrier exponentially suppresses the wavefunction overlap and interlayer tunneling (Extended Data Fig. 6). Nor can it result from magnetic dipolar coupling, which is merely of the order $\frac{\mu_0}{4\pi} \frac{\mu_B^2}{d^3} = 7$ neV ($\mu_0$: vacuum permeability). Moreover, phase II$_A$ strengthens with density, inconsistent with single-exciton mechanisms because exciton binding energy decreases with density (Extended Data Fig. 4). Nor can a single-particle picture explain the behaviors in phases II$_B$ and I.

Having excluded single-particle origins, the three different phases must stem from collective many-body correlations in a quantum exciton fluid. Our excitons lie deep in the quantum-degenerate regime. Both renormalization group and Monte Carlo simulations give a BKT transition temperature in the dilute limit as[41–43,12,13]

$$k_B T_{\text{BKT}} \approx 1.3 \frac{\hbar^2}{m_x} n_x \qquad (1)$$

where $m_x \approx m_0$ is the exciton mass ($m_0$: bare electron mass). A typical $n_x = 0.5 \times 10^{12} \text{cm}^{-2}$ gives $T_{\text{BKT}} \approx 6$ K, far exceeding our milli-Kelvin base temperature. We therefore expect that exciton condensation will become important.

## Two-component exciton condensate

We now show that a strongly interacting two-component exciton condensate can account for all our observations. Although four exciton flavors exist, previous theoretical studies show that the interacting excitons can have only one or two flavors condensing simultaneously[13,14,44]. Their collective behavior is captured by an interacting boson model that phenomenologically describes multicomponent interlayer excitons, incorporating the TMD valley structure and fermionic e-h interactions at the mean-field level[13]. The Hamiltonian is

$$H = \sum_i (E_i - \mu) n_i + \frac{g_H + g_X}{2} \left( \sum_i n_i \right)^2 - g_X (n_1 n_2 + n_3 n_4) \qquad (2)$$

where $n_i = \psi_i^\dagger \psi_i$, and $\psi_i$ is the bosonic field operator for exciton flavor $i$ (1=KK, 2=K'K', 3=KK', 4=K'K; first and second labels are electron and hole valleys, respectively). The chemical potential $\mu$ controls the total exciton density. The interaction constants $g_H = 8\pi d$ (in atomic units with Bohr radius $a_B = 1.5$ nm, Rydberg energy Ry $= 67$ meV) and $g_X$ describe the Hartree and exchange interactions between excitons, respectively[13]. Momentum dependence is neglected under the assumption of a spatially uniform condensate. Due to different electron and hole spin combinations, the flavor-dependent single-exciton energy $E_i$ in a magnetic field has the form

$$E_i = \begin{cases} (-g_c + g_v)\mu_B B - \Delta, & i = 1 \text{ (KK)} \\ (+g_c - g_v)\mu_B B - \Delta, & i = 2 \text{ (K'K')} \\ (-g_c - g_v)\mu_B B, & i = 3 \text{ (KK')} \\ (+g_c + g_v)\mu_B B, & i = 4 \text{ (K'K)} \end{cases} \quad (3)$$

Fig. 3a schematically illustrates these energy levels. At $B = 0$, time-reversal symmetry ensures degeneracy within the two intravalley flavors (blue lines) and two intervalley flavors (yellow), while a small energy splitting $\Delta$ between the two groups is allowed. Under a magnetic field, intervalley excitons shift faster than intravalley ones because the electron and hole Zeeman contributions add for the former but subtract for the latter.

Most parameters in the Hamiltonian can be reliably estimated experimentally (Methods); however, the exchange interaction $g_X$ and the zero-field splitting $\Delta$ are difficult to determine for strongly interacting excitons and are treated phenomenologically. From the interacting boson Hamiltonian, it is clear that the sign of $g_X$ determines the exciton flavor-flavor interaction, and therefore condensate symmetry. A positive $g_X$ creates a two-component condensate with equal density in two condensed flavors, whereas a negative $g_X$ leads to a ferromagnetic single-component condensate that fully polarizes at an infinitesimal field (Extended Data Fig. 7). Hartree-Fock calculations predict $g_X < 0$ and thus a ferromagnetic exciton condensate at our interlayer spacing[13], in clear contradiction with our MCD observations. This is likely due to Hartree-Fock's well-known tendency to overestimate ferromagnetism. To capture our experimental behavior, we model the interacting boson system using phenomenological parameters with $g_X = 1$ (in atomic units) and $\Delta = 1$ µeV (Methods).

By minimizing the Hamiltonian, we obtain the condensate order parameters for all four exciton flavors. Fig. 3c plots the calculated spin-valley polarizations at zero temperature. The calculation successfully captures the reversed electron susceptibility and reduced hole susceptibility in phase II$_A$, abrupt polarization jumps at the II$_A$/II$_B$ boundary, and the fully polarized phase I emerging above the tesla-scale saturation field. The same set of parameters also captures the observed density dependence, as shown in the calculated electron spin susceptibility in Fig. 3d. The model identifies phase II$_A$ as an intravalley two-component condensate (KK+K'K'), phase II$_B$ as an intervalley two-component condensate (KK'+K'K), and phase I as a single-component condensate. Their relative energetics are illustrated in Fig. 3b, with quantitative details provided in Extended Data Fig. 7.

At zero field, the two-component states have lower energy because a positive $g_X$ favors equal occupation of two flavors to minimize the exchange energy. A positive $\Delta$ further selects the intravalley two-component condensate (phase II$_A$), in which the KK and K'K' order parameters have equal magnitude and a fixed relative phase (Fig. 3e). A small $B$ field slightly increases the K'K' condensate density (Fig. 3f), but the resulting imbalance is strongly suppressed by the large exchange-induced stiffness. Since only intravalley flavors condense, the electron and hole valley polarizations are locked together, both equal to $n_2 - n_1 = 2(g_v - g_c)\mu_B B/g_X$. This locking forces the electron to polarize against the field while the hole polarizes with the field ($g_v > g_c$), resulting in the reversed eMCD slope in phase II$_A$. In an ideal system at zero temperature, the electron and hole polarizations should be equal in magnitude. However, any normal fluid component, either due to thermal excitation or sample inhomogeneity, will add a positive background, reducing the magnitude of the reversed electron polarization.

With increasing field, the intervalley two-component condensate (II$_B$) acquires a larger quadratic Zeeman energy lowering than II$_A$ owing to its larger g-factor, as illustrated by the blue and yellow curves in Fig. 3b. A first-order quantum phase transition occurs at their energy crossing point. The exciton condensate order parameter suddenly switches from two intravalley flavors to two intervalley flavors, producing the observed polarization jumps at the II$_A$/II$_B$ boundary. The relatively soft experimental transition likely reflects finite temperature or inhomogeneity broadening. In phase II$_B$, the exchange interaction still resists flavor imbalance (Fig. 3g), but the larger effective g-factor $g_c + g_v$ induces a faster polarization than phase II$_A$. The intervalley character enforces opposite electron and hole valley polarizations of equal magnitude $n_3 - n_4 = 2(g_c + g_v)\mu_B B/g_X$, exactly as seen experimentally.

At an even stronger field, the linear Zeeman lowering of KK' exciton energy (pink line in Fig. 3b) dominates, and the system evolves into a fully polarized single-component condensate (phase I, Fig. 3h). The saturation field (~1 T) is set by the meV-scale inter-flavor exchange energy.

## Exciton condensate phase diagram at finite temperatures

Having established that the unusual spin behavior originates from two-component BEC, we now examine the temperature dependence of the exciton condensate. Fig. 4a shows the electron spin susceptibility (more precisely, d(eMCD)/d$B$) as a function of $B$ and $T$ at a fixed exciton density $0.5 \times 10^{12}$ cm$^{-2}$. The blue-white region at low $B$ and low $T$ corresponds to suppressed (and slightly reversed) spin susceptibility from the II$_A$ condensate, and the strongly red region at higher $B$ and low $T$ corresponds to enhanced spin susceptibility from the II$_B$ condensate. Both features gradually fade with increasing temperature and vanish once the exciton fluid becomes a paramagnetic normal state. Fig. 4b shows the average susceptibilities in two different field windows, $|B| < 30$ mT (II$_A$, blue) and $30 < B < 70$ mT (II$_B$, yellow). Both curves exhibit a sudden change in the temperature derivative at 1.6 K, identifying the BKT transition temperature $T_{\text{BKT}}$. A common $T_{\text{BKT}}$ for both II$_A$ and II$_B$ phases follows from the identical masses and exciton densities, and from the fact that the flavor imbalance at tens-of-mT fields is too small to appreciably alter the superfluid stiffness. Above 1.6 K the two curves coincide with no difference (Fig. 4b, lower panel), consistent with a paramagnetic, non-condensed state. Below 1.6 K, the susceptibility decreases in II$_A$ and increases in II$_B$, reflecting the growth of the respective superfluid fractions upon cooling.

We can use the difference between the two curves as a proxy of superfluid density, since the normal-fluid component is paramagnetic and doesn't contribute to the difference. Fig. 4c displays this difference as a function of $n_x$ and $T$. A clear condensate dome emerges, bounded on the high-density side by quantum melting at the Mott transition and on the high-temperature side by the BKT transition. The theoretical BKT temperature in Eq. (1) assumes a single-flavor dilute gas; for a two-component condensate, it is reduced by roughly half[45]. The blue dashed line in Fig. 4c corresponds to $k_B T_{BKT} = 0.65 \frac{\hbar^2}{m_x} n_x$. It matches the exciton condensate phase boundary well in the dilute limit ($< 0.2 \times 10^{12} \text{cm}^{-2}$). At higher densities, the measured transition temperature shows a sublinear density dependence, likely due to strong many-body interactions[46]. The maximum achievable BKT temperature is around 1.8 K near the Mott transition. Individual excitons are known to remain well bound until nearly 100 K (refs.[6,7]). The transition therefore marks a loss of phase coherence rather than pair breaking.

Finally, we examine how the condensate varies with device geometry. Similar condensate features are observed in other TMD combinations – for example, electron-doped WSe$_2$ paired with hole-doped MoSe$_2$ (Extended Data Fig. 8) – indicating a generic multicomponent condensate in e-h bilayers. The remaining device parameter is the TMD twist angle $\theta$. We found devices with near-60° twist angles exhibit all three condensate phases II$_A$, II$_B$ and I at different magnetic fields. In contrast, phase II$_A$ becomes progressively weaker as $\theta$ deviates from 60°, and disappears near 0° twist angle, where only II$_B$ and I persist (Extended Data Fig. 1). Multi-region devices confirm that this strong angle dependence is intrinsic rather than uncontrolled device-to-device variations (Extended Data Fig. 2-3). This trend could be understood as an angle-dependent $\Delta$ splitting that is optimal for intravalley exciton condensates at $\theta \approx 60°$, although its microscopic origin remains to be investigated.

In summary, our results establish that dipolar excitons in MoSe$_2$/hBN/WSe$_2$ e-h bilayers realize a strongly interacting two-component exciton condensate. It provides a platform for exploring high-temperature spinor BEC with non-trivial internal degrees of freedom. Although modest in absolute terms, the transition temperature is already much higher than other multicomponent BECs. Furthermore, even higher transition temperature is possible in systems with higher exciton Mott density, which can be achieved by new van der Waals e-h bilayer systems with closer interlayer spacing, smaller dielectric constant, and/or larger effective mass.

## Methods

**Device fabrication.** We assemble the TMD heterostructures using a dry-transfer technique employing polyethylene-terephthalate-glycol (PETG) stamps. Monolayer MoSe$_2$ and WSe$_2$, few-layer graphene, and hBN flakes are mechanically exfoliated from bulk crystals onto SiO$_2$/Si substrates. Few-layer graphene is used as both the top and bottom gates; 5-10 nm hBN serves as the gate dielectric; and a 1-3 nm hBN flake is employed as the interlayer spacer. Electrical contacts to MoSe$_2$ are made using few-layer graphene, whereas contacts to WSe$_2$ are either platinum (devices D4-D7) or graphene (other devices). A 0.5-mm-thick PETG stamp is used to pick up the layers at 65-75 °C. A >100-nm-thick sacrificial hBN flake is first picked up to protect subsequent layers. All subsequent layers are then sequentially stacked and finally released onto a high-resistivity SiO$_2$/Si substrate at 100 °C. After transfer, the PETG film is dissolved in chloroform. Metal contacts (5 nm Cr / 50-80 nm Au) are patterned by photolithography or e-beam lithography and deposited by e-beam evaporation. For devices D2, D5 and D8, the sacrificial hBN is eventually removed by XeF$_2$ etching.

To control the twist angle, TMD flakes with straight cleavage edges are selected. The nominal angle (modulo 60°) is designed using these edges and subsequently verified by polarization-resolved second-harmonic-generation (SHG) measurements performed on the completed devices using a 900 nm femtosecond laser. The SHG response from the heterostructure region is compared with that of individual monolayers. Increased (decreased) SHG signal in the heterostructure region indicates constructive (destructive) interference from a 0° (60°) alignment. To ensure that the observed twist-angle dependence of MCD is intrinsic rather than arising from uncontrolled device-to-device variations (e.g., defect density, asymmetric gate screening, or contact quality), we fabricated two multi-region devices (D2 and D3). Each device contains several regions with different $\theta$ values but made from the same exfoliated flakes and with identical gate, dielectric, and contact configurations (Extended Data Figs. 2-3).

**Optical measurements.** All optical measurements are performed in a dilution refrigerator (Bluefors LD250) with a base lattice temperature of 10 mK. Signal wires are heavily filtered at the mixing chamber flange using a series combination of QDevil QFilters and Basel Precision Instruments MFT25 filters. Gate and bias voltages on the device are applied with Keithley 2400/2450 source meters or Keithley 2502 picoammeters.

The light source for reflection spectroscopy and MCD measurements is either a diode laser driven below lasing threshold or a supercontinuum source. The reflected light from the sample is dispersed by a spectrometer (Andor Shamrock 303i) and collected by a CCD camera (Andor iDus DU420A-BEX2-DD). For MCD measurements, we use a small incident laser power (< 1 nW) to ensure minimal heating effects. The electron temperature under such laser power is estimated to be about 200 mK. Using even weaker laser power would not substantially lower the electron temperature further, because the interlayer tunneling current (Extended Data Fig. 6) already generates a sub-nanowatt Joule heating. For reflectance measurements, higher power (a few nW) is sometimes used. Electron and hole densities are extracted from reflectance spectra following the procedure described in ref.[7].

The MCD spectrum is defined as $\frac{R_+ - R_-}{R_+ + R_-}$, where $R_\pm$ denotes the reflectance for the two circular polarizations. The raw MCD data are antisymmetrized with respect to $\pm B$. For each TMD layer, the MCD spectrum is integrated over a narrow wavelength window centered at the attractive polaron (intralayer trion) resonance. The integrated signal in another wavelength window, placed at a featureless but nearby wavelength, is subtracted as the background to eliminate any residual MCD from the optical path.

**Interacting boson model**. The interacting boson model calculations follow ref.[13] with several modifications tailored to the present system. Starting with the mean-field boson Hamiltonian derived in ref.[13], we introduce the flavor-dependent energy shift $E_{cv}$ defined in Eq. (3), and the Hamiltonian reads

$$H = \sum_{c,v}(E_{cv} - \mu)\,\psi_{cv}^\dagger \psi_{cv} + \frac{1}{2} \sum_{c,c',v,v'} \left[ g_\text{H} \psi_{cv}^\dagger \psi_{c'v'}^\dagger \psi_{c'v'} \psi_{cv} + g_\text{X} \psi_{cv}^\dagger \psi_{c'v'}^\dagger \psi_{cv'} \psi_{c'v} \right],$$

where conduction and valence band valley indices $\{c, v, c', v'\}$ run over K and K'. Upon explicitly expanding the sums, the Hamiltonian can be reorganized into Eq. (2) in terms of the flavor densities

$n_i = \psi_i^\dagger \psi_i$ at the mean-field level. We numerically minimize the Hamiltonian with respect to $\{n_i\}$ to obtain the ground state energy and condensate density. It is then straightforward to obtain the electron spin-valley polarization $(n_1 + n_3 - n_2 - n_4)/(n_1 + n_2 + n_3 + n_4)$ and hole polarization $(n_1 + n_4 - n_2 - n_3)/(n_1 + n_2 + n_3 + n_4)$.

The calculation parameters are taken as follows. The conduction and valence g-factors are $g_c = 3$ and $g_v = 6$, respectively[39,40]. With the effective electron mass $m_e = 0.6 m_0$, hole mass $m_h = 0.4 m_0$, and dielectric constant $\epsilon = 7$, the Bohr radius is $a_B = \epsilon \hbar^2/[e^2 m_e m_h/(m_e + m_h)]$ =1.543 nm, and the Rydberg energy is $Ry = e^2/(2\epsilon a_B) = 66.64$ meV. We take $d = 1$ in the calculations (corresponding to 4-5 layers of hBN). Two parameters cannot be reliably obtained from first principles: the exchange interaction $g_X$ and the splitting $\Delta$. In principle $g_X$ can be calculated using Hartree-Fock theory, but such calculations are unreliable because the sign of $g_X$ is determined by the small difference between two competing terms. The results have the wrong sign compared to the experiment[13]. We therefore choose $g_X = 1$ to approximately match the experimentally observed saturation field. The $\Delta$ splitting is subtle and can originate from trigonal warping and Berry phase effects[47], which are hard to model at finite density. We therefore take $\Delta = 1$ µeV to match the experimentally observed phase II$_A$ width. Extended Data Fig. 7 shows the calculation results for other values. Taking $\Delta = -1$ µeV makes phase II$_B$ favorable even at small $B$ (Extended Data Fig. 7d-f), which can describe the behavior of near 0° devices. Taking a negative $g_X = -1$ will result in a single-flavor condensate regardless of $\Delta$, and both electrons and holes become fully polarized at infinitesimal field (Extended Data Fig. 7g-i).

The last term in Eq. (2) immediately shows that no more than two flavors condense simultaneously. We can therefore get a more intuitive understanding by restricting to a two-flavor subspace, e.g., $n_1$ and $n_2$ with $n_3 = n_4 = 0$. The Hamiltonian then reduces to a simple Ginzburg-Landau form

$$H = -(\mu + \Delta)(n_1 + n_2) + \frac{1}{2}(g_H + g_X)(n_1 + n_2)^2 - g_X n_1 n_2 - (g_v - g_c)\mu_B B(n_1 - n_2)$$

The stationary condition $\frac{\partial H}{\partial (n_1 + n_2)} = 0, \frac{\partial H}{\partial (n_1 - n_2)} = 0$ gives a simple linear result

$$n_1 + n_2 = \frac{\mu + \Delta}{g_H + g_X/2}, \quad n_1 - n_2 = \frac{2(g_v - g_c)\mu_B B}{g_X}$$

This gives the analytic polarization in phase II$_A$; the result for phase II$_B$ follows by exchanging $(g_v - g_c) \rightarrow (g_v + g_c)$ and $\mu + \Delta \rightarrow \mu$. These expressions agree exactly with the full numerical minimization of Eq. (2).

## Data availability

The data that support the findings of this study are available from the corresponding author upon request.

## Acknowledgements

We thank Dr. Michael Zaletel, Dr. Tianle Wang and Dr. Zhihuan Dong for insightful discussions. This work was primarily funded by the U.S. Department of Energy, Office of Science, Basic Energy Sciences, Materials Sciences and Engineering Division under Contract No. DE-AC02-05-CH11231 within the van der Waals heterostructure program KCFW16 (device fabrication, dilution refrigerator MCD measurements). Reflection spectroscopy of the excitonic insulator was supported by the AFOSR award FA9550-23-1-0246. K.W. and T.T. acknowledge support from the JSPS KAKENHI (Grant Numbers 21H05233 and 23H02052), the CREST (JPMJCR24A5), JST and World Premier International Research Center Initiative (WPI), MEXT, Japan. R.Q. and H.K. acknowledge support from Kavli ENSI graduate student fellowship.

## Author contributions

F.W. and R.Q. conceived the project. K.W. and T.T. synthesized the hBN crystals. Q.L., R.Q., R.X. and J.N. fabricated the devices. H.L. and J.X. assisted with twist-angle determination. R.Q., H.K. and Q.L. set up and carried out the optical measurements. R.Q. performed the theoretical calculations. R.Q., Q.L., J.N. and F.W. analyzed the data with input from M.F.C. and A.H.M. All authors discussed the results and contributed to writing the manuscript.

## Competing interests

The authors declare no competing interests.

# Figures

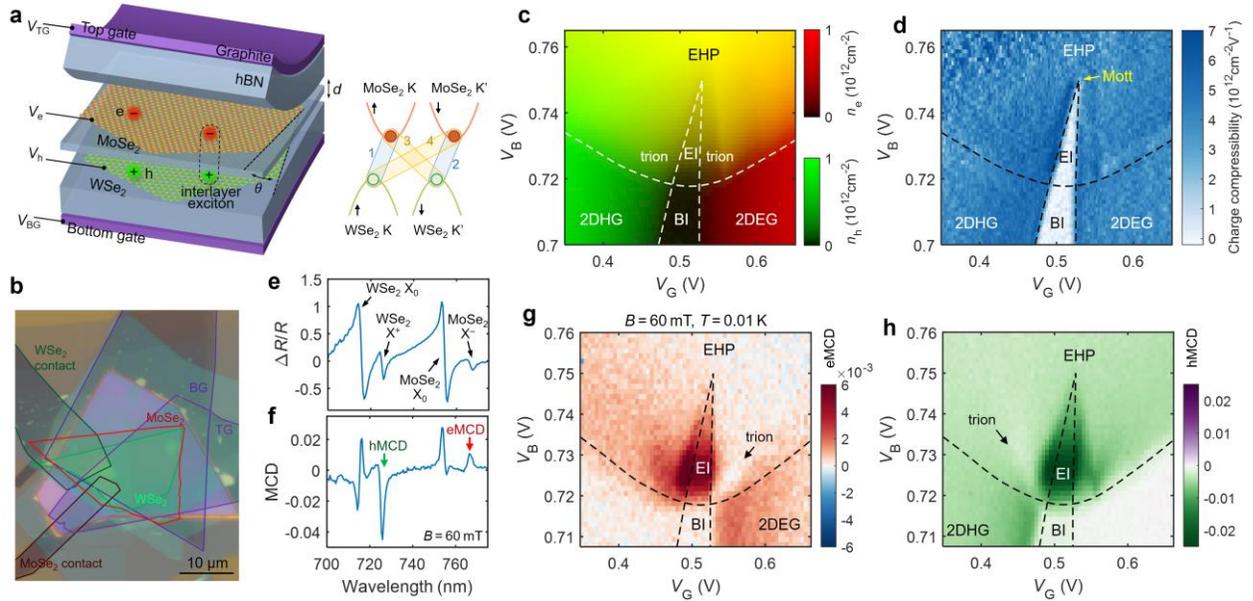

**Fig. 1 | Equilibrium exciton fluids in electron-hole bilayers.**

**a**, Schematic of the MoSe$_2$/hBN/WSe$_2$ e-h bilayer. Inset: illustration of four possible interlayer exciton flavors formed from electrons and holes in the K and K′ valleys. Black arrows indicate the spin of conduction and valence electrons (hole spin is opposite).

**b**, Optical micrograph of device D1 (5-layer hBN spacer, near 60° angle alignment).

**c**, Experimentally determined electron density $n_e$ in MoSe$_2$ (red channel) and hole density $n_h$ in WSe$_2$ (green channel) as functions of $V_G$ and $V_B$. The vertical electric field is fixed by antisymmetric gating $V_{TG} - V_{BG} = 6.6$ V, and the MoSe$_2$ layer is grounded. The phase diagram includes a trivial BI (black), electron or hole Fermi liquids (2DEG, red; 2DHG, green), and strongly interacting e-h fluids (yellow). At charge neutrality the e-h fluids form an EI.

**d**, Bilayer charge compressibility[7] obtained from the density data in **c**. Incompressible regions (white) are gapped insulating states, including the BI with no charge carriers, and the EI at finite and equal e-h densities.

**e,** Representative reflection contrast $\Delta R/R$ spectrum at an exciton doping $0.3 \times 10^{12}$ cm$^{-2}$. Two characteristic intralayer optical resonances (X$_0$ and X$^-$/X$^+$) from each TMD layer are observed.

**f**, Corresponding MCD spectrum at $B = 60$ mT. The MCD intensity at MoSe$_2$ X$^-$ (WSe$_2$ X$^+$) wavelength, labeled as eMCD (hMCD), is proportional to electron (hole) spin-valley polarization.

**g**, MCD signal from the MoSe$_2$ X$^-$ resonance (red arrow in **f**; eMCD for short) as a function of applied voltages, measured at 60 mT. The signal is approximately proportional to the electron spin-valley polarization (positive values indicate polarization toward the K valley).

**h**, MCD signal from the WSe$_2$ X$^+$ resonance (hMCD; green arrow in **f**), which is roughly proportional to the hole polarization. All data are taken at $T = 0.01$ K.

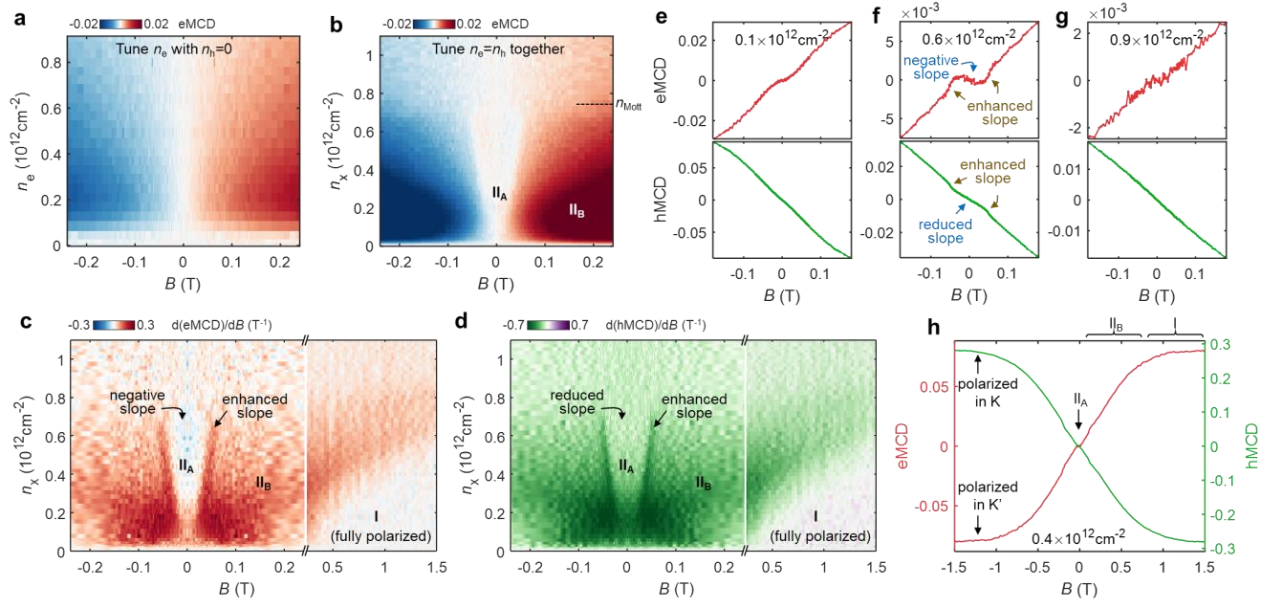

**Fig. 2 | Spin-valley polarization in the exciton fluid.**

**a**, Measured eMCD in the 2DEG phase, as a function of $n_e$ and $B$. The applied $V_B$ is below the bandgap so that $n_h = 0$. The linear dependence of eMCD signal on $B$ is consistent with the Pauli paramagnetism of a Fermi liquid.

**b**, eMCD measured along the net charge neutral condition as a function of pair density $n_x$ ($= n_e = n_h$) and $B$. A nearly white region centered at zero field exhibits strongly suppressed eMCD.

**c**, $B$-derivative of the eMCD data shown in **b**, corresponding to the electron spin-valley susceptibility. Three distinct phases are recognized: phase II$_A$ near zero field has suppressed or even negative susceptibility; phase II$_B$ at moderate field exhibits large positive susceptibility; phase I at high field is fully polarized with zero susceptibility. A different horizontal scale is used after the axis break at 0.24T.

**d**, $B$-derivative of the associated hMCD data, showing the hole susceptibility. It also exhibits three distinct phases similar to the eMCD behavior in **c**.

**e-g**, Representative line plots of eMCD and hMCD for exciton densities at 0.1, 0.6 and $0.9 \times 10^{12}$cm$^{-2}$, respectively. The two lower densities lie in the EI phase, while the highest density is beyond $n_{Mott} \approx 0.75 \times 10^{12}$cm$^{-2}$ and in the EHP phase. The II$_A$ phase of the exciton fluid is most prominent at $n_x = 0.6 \times 10^{12}$cm$^{-2}$, where eMCD develops a slightly negative slope (corresponding to electron spins polarizing against the field), and hMCD exhibits a strongly suppressed slope at low magnetic field.

**h**, eMCD (left axis) and hMCD (right axis) at exciton density $0.4\times 10^{12}\,\mathrm{cm}^{-2}$ over an extended field range. In phases $II_B$ and $I$, the electron and hole valley polarizations are equal in magnitude and opposite in sign, and reach full polarization at the same saturation field.

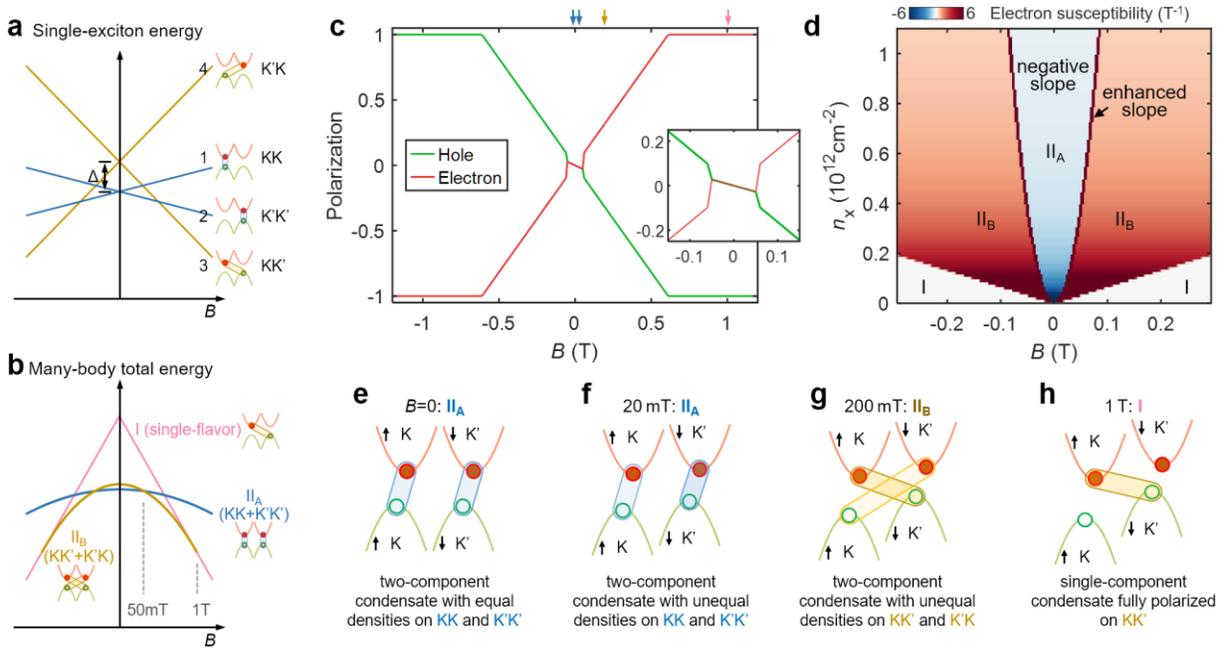

**Fig. 3 | Two-component condensates and quantum phase transitions.**

**a**, Schematic single-exciton energy diagram for four exciton flavors under a magnetic field (see Eq. (3)). Their Zeeman shift differs due to different combinations of electron and hole g-factors. A zero-field splitting $\Delta$ between intravalley and intervalley species is allowed by symmetry.

**b**, Many-body total energy for three competing condensate states: intravalley two-component condensate (phase $II_A$, blue), intervalley two-component condensate (phase $II_B$, yellow), and fully-polarized single-component condensate (phase $I$, pink).

**c**, Electron (red) and hole (green) polarizations calculated by the interacting boson model. Inset, a zoom-in view near zero field. The calculation reproduces almost all observations in Fig. 2.

**d**, Calculated electron susceptibility ($B$-derivative of electron spin-valley polarization) at different exciton densities. The three exciton condensate phases agree well with the experimental data in Fig. 2.

**e-h**, Schematic pairing order parameter in different condensate phases. Phase $II_A$ at low field (**e**, **f**) is an intravalley two-component condensate. Phase $II_B$ at moderate field (**g**) is an intervalley two-component condensate. Phase $I$ at high field (**h**) is a fully polarized single-component condensate.

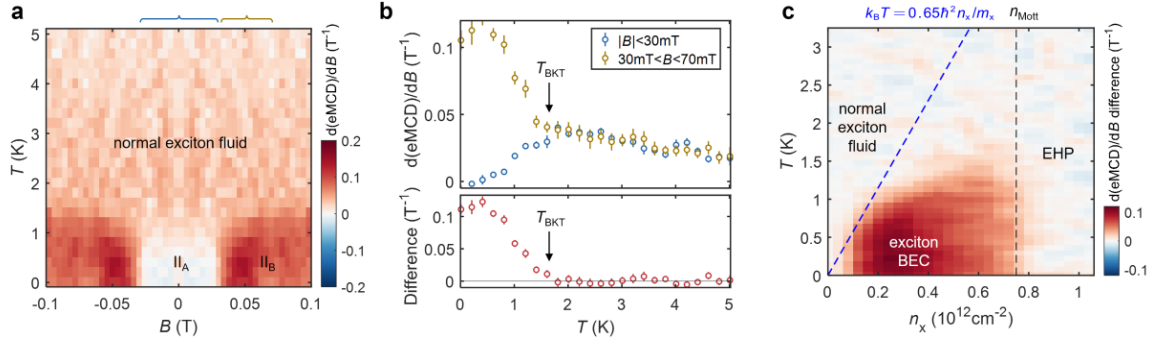

**Fig. 4 | Condensate phase diagram.**

**a**, $B$-derivative of eMCD as a function of $B$ and $T$, at $n_x = 0.5 \times 10^{12}\,\text{cm}^{-2}$. The blue and yellow brackets mark the two field windows used for averaging in **b**.

**b**, Temperature dependence of d(eMCD)/d$B$, averaged in two different $B$ windows, $|B| < 30$ mT (corresponding to phase II$_A$, blue) and $30\,\text{mT} < B < 70\,\text{mT}$ (phase II$_B$, yellow). Their difference (lower panel) is zero at high temperatures, but increases quickly below $T_{\text{BKT}} \approx 1.6$ K.

**c**, The difference between d(eMCD)/d$B$ within $30\,\text{mT} < B < 70\,\text{mT}$ and d(eMCD)/d$B$ within $|B| < 30$ mT as a function of pair density and temperature. A condensate dome is bounded by the Mott transition at high density and by the BKT transition at high temperature. The blue dashed line shows $k_B T = 0.65 \dfrac{\hbar^2}{m_x} n_x$, the expected BKT temperature for weakly interacting two-component BEC.

# Extended Data Figures

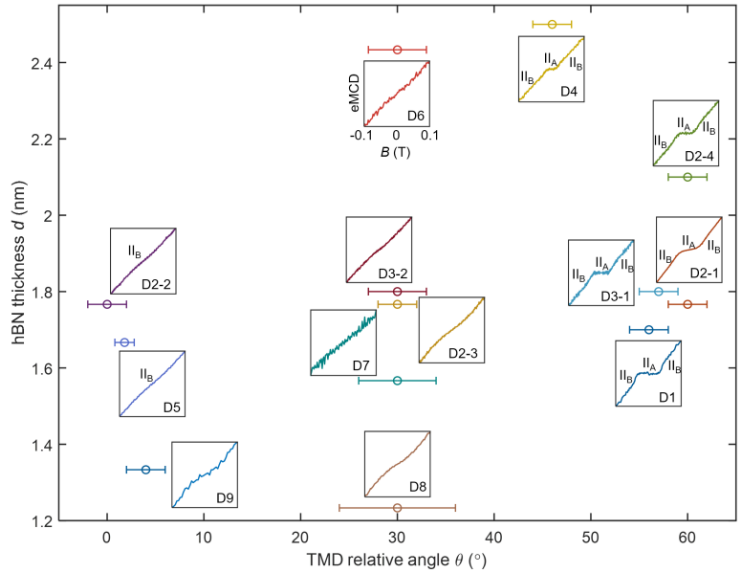

**Extended Data Fig. 1 | Summary of eMCD results for multiple devices.**

Evolution of condensate phases with interlayer distance $d$ and MoSe$_2$-WSe$_2$ relative rotation angle $\theta$. Each subpanel plots the eMCD (vertical axis) versus $B$ (horizontal axis, all in range $-0.1$ to $0.1$ T) for one device, measured at exciton density $0.4 \times 10^{12}\,\text{cm}^{-2}$. For ~60° devices, the two-component condensates exhibit the first order phase transition between II$_A$ and II$_B$ states. The transition critical field decreases with increasing interlayer spacing. For ~0° devices, no such transition is observed, and the condensate stays in II$_B$ even at zero field.

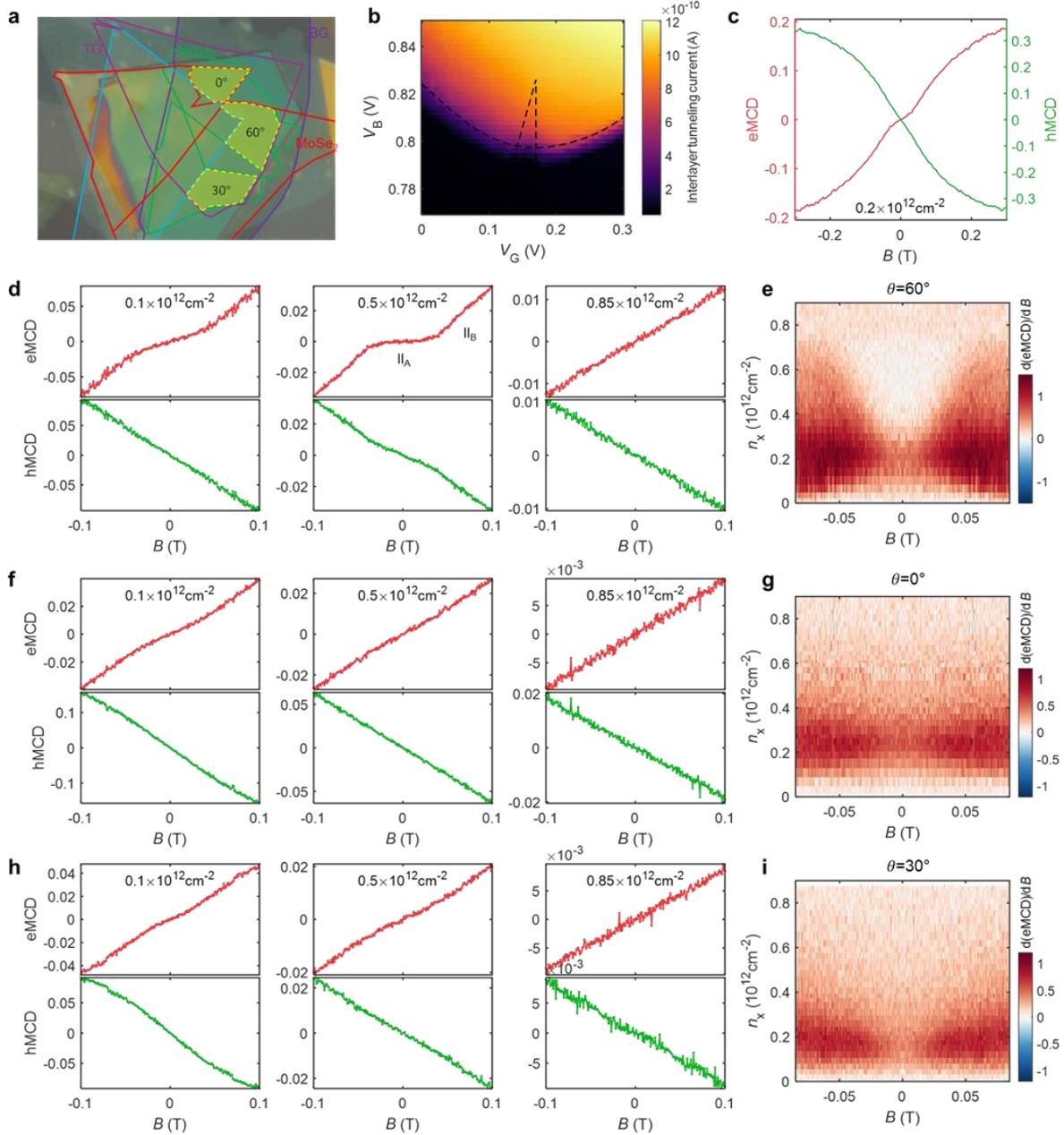

**Extended Data Fig. 2 | Main results for device D2.**

**a**, An optical image of device D2. In this device, the same TMD flakes are made into different regions with relative angle 0°, 30° and 60°.

**b**, Interlayer tunneling current, which is still on nanoampere level but larger than device D1. The increased tunneling generates more heating, which likely causes a higher electron temperature for this device.

**c**, eMCD (left axis) and hMCD (right axis) at exciton density $n_x = 0.2 \times 10^{12} cm^{-2}$, measured in the 60° region.

**d**, eMCD and hMCD at pair densities 0.1, 0.5 and $0.85\times 10^{12}\,\mathrm{cm}^{-2}$ for the 60° region.

**e**, Derivative of eMCD along the charge neutral line for the 60° region.

**f-g**, Same data but measured in the 0° region. Phase $II_A$ is almost unrecognizable.

**h-i**, Same data but measured in the 30° region. Phase $II_A$ is very weak.

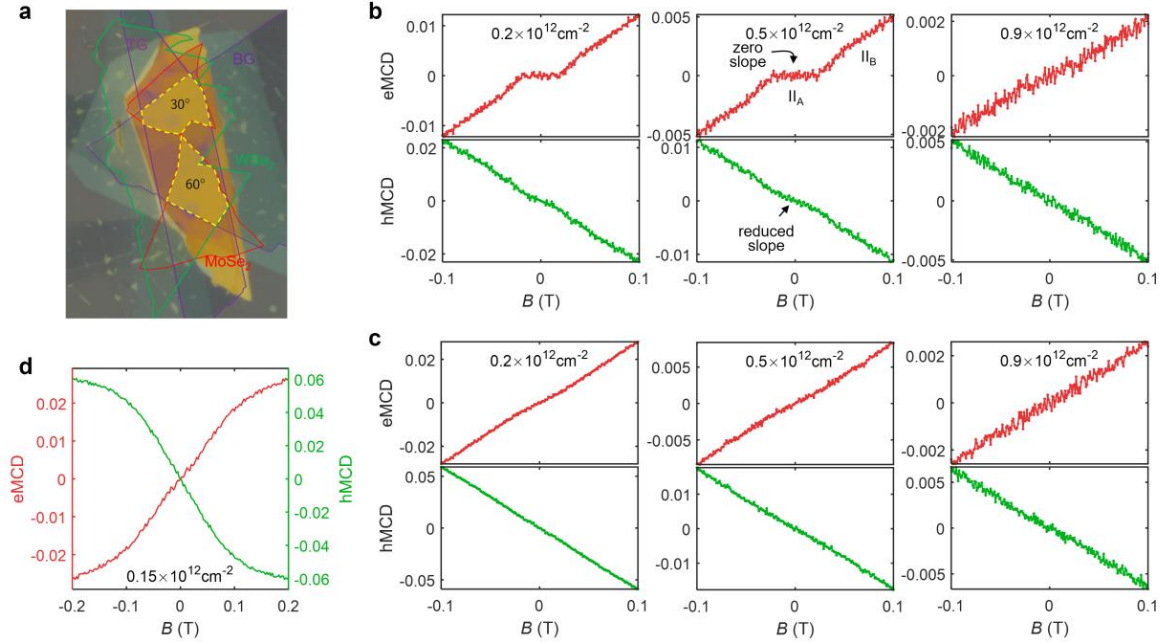

**Extended Data Fig. 3 | Main results for device D3.**

**a**, An optical image of device D3. In this device, the same TMD flakes are made into two different regions with relative angle 30° and 60°.

**b-c**, eMCD and hMCD at pair densities 0.1, 0.5 and $0.9\times 10^{12}\,\mathrm{cm}^{-2}$, measured in the 60° region (**b**) and the 30° region (**c**) respectively.

**d**, eMCD (left axis) and hMCD (right axis) at $n_x = 0.15 \times 10^{12}\,\mathrm{cm}^{-2}$ for the 30° region. Although the phase $II_A$ is not well developed for this twist angle, on a larger field range electrons and holes still polarize with the same magnitude and opposite sign, saturating at the same field. This suggests that we are still in the condensate regime with the same phase $II_B$ and phase I physics.

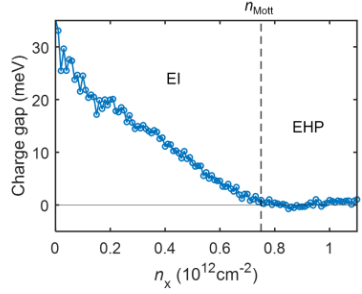

**Extended Data Fig. 4 | Density-dependent charge gap and exciton Mott transition.**

The charge gap as a function of pair density (device D1), extracted using the method described in ref.[7]. In the dilute limit, the charge gap (exciton binding energy) is about 30 meV. It continuously decreases to zero at the Mott density $n_{Mott} \approx 0.75 \times 10^{12} cm^{-2}$, beyond which the system becomes gapless. After this point, $n_x$ should be interpreted as the e-h pair density rather than the exciton density.

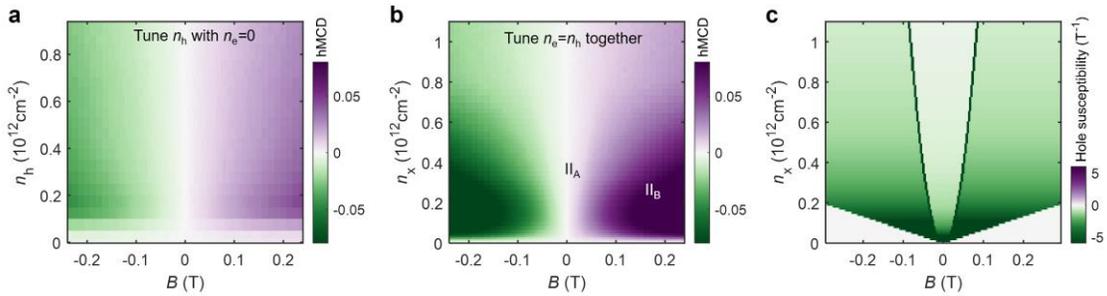

**Extended Data Fig. 5 | Additional data from the hole side.**

**a-b**, Comparison of density- and field-dependent hMCD signal in the 2DHG phase (**a**) and along the charge neutral line (**b**). The 2DHG exhibits linear polarization; the EI phase has a reduced signal in II$_A$.

**c**, Calculated hole susceptibility as a function of $B$ and $n_x$. The reduced but not reversed susceptibility in II$_A$, the abrupt polarization jump at II$_A$/II$_B$ boundary, and the fully polarized phase I all match well with the experimental data in Fig. 2.

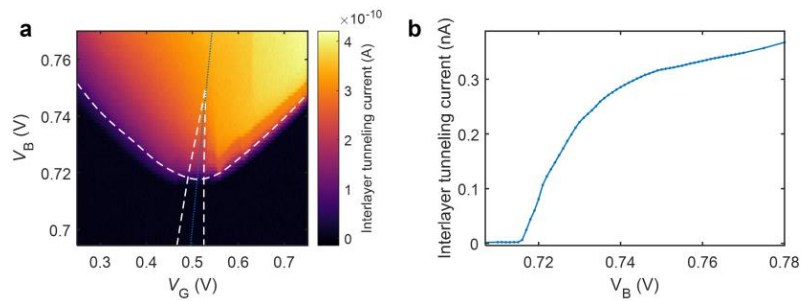

**Extended Data Fig. 6 | Interlayer tunneling.**

**a**, Interlayer tunneling current as a function of $V_G$ and $V_B$ (device D1).

**b**, Tunneling current versus $V_B$ along the charge neutrality line (blue dotted line in **a**). The tunneling lifetime can be estimated to be at least 0.5 ms.

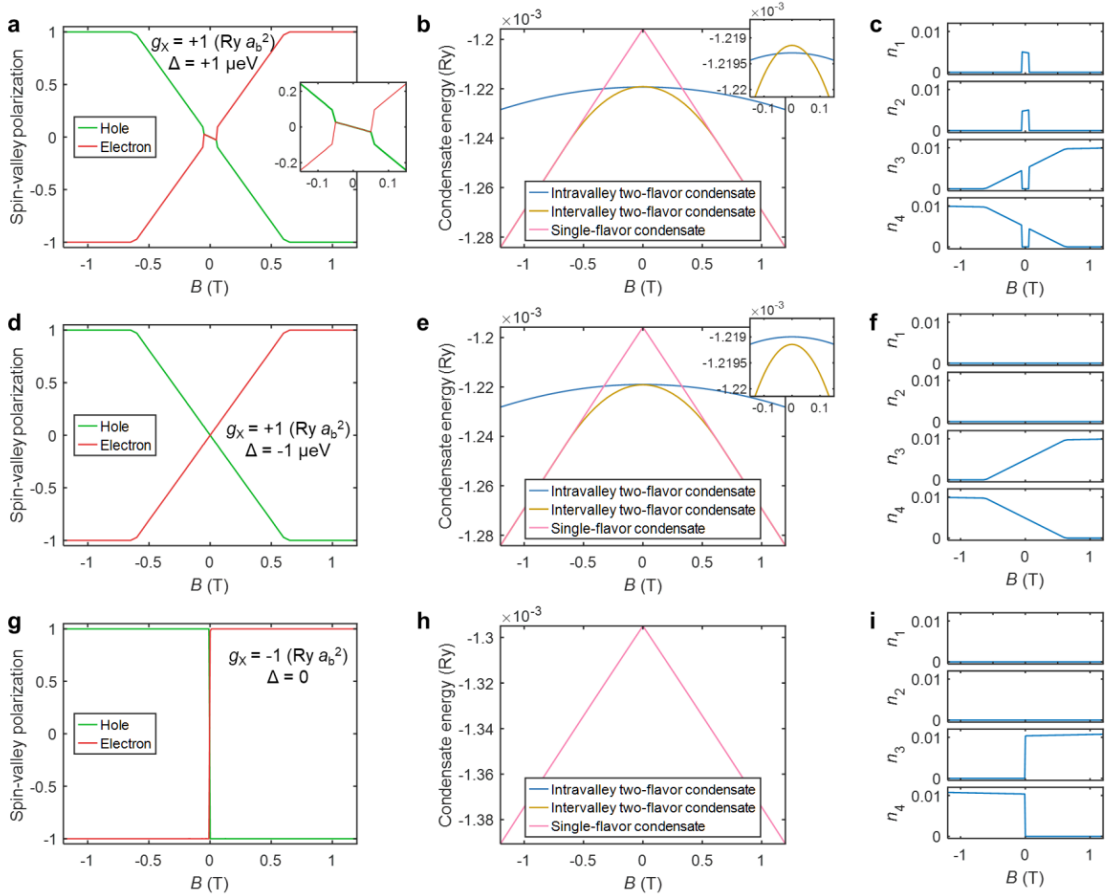

**Extended Data Fig. 7 | Additional calculation results.**

**a**, Calculated electron (red) and hole (green) spin-valley polarizations with a positive $g_X = 1$ (in atomic units) and a positive splitting $\Delta = 1$ µeV favoring intravalley flavors (parameters used in the main text). Inset, a zoom-in plot near zero field.

**b**, Calculated condensate total energy per area for three competing condensate states: intravalley two-component condensate ($II_A$, blue), intervalley two-component condensate ($II_B$, yellow), and single-component condensate (I, pink). Inset, a zoom-in plot near zero field clearly showing the energy crossing between $II_A$ and $II_B$.

**c**, Condensate density in four exciton flavors. The order parameters suddenly switch from two intravalley flavors to two intervalley flavors at the energy crossing at ~50 mT.

**d-f**, Same plots but with a negative $\Delta = -1$ µeV. Now phase $II_B$ always has lower energy than $II_A$ even at zero field. This can explain the absence of MCD suppression in near 0° devices.

**g-i**, Same plots but with a negative $g_X = -1$. With attractive interactions, two-component condensates are no longer local energy minima, so their energy cannot be calculated in **h**. In such single-component condensates, any infinitesimal field will lead to full valley polarization.

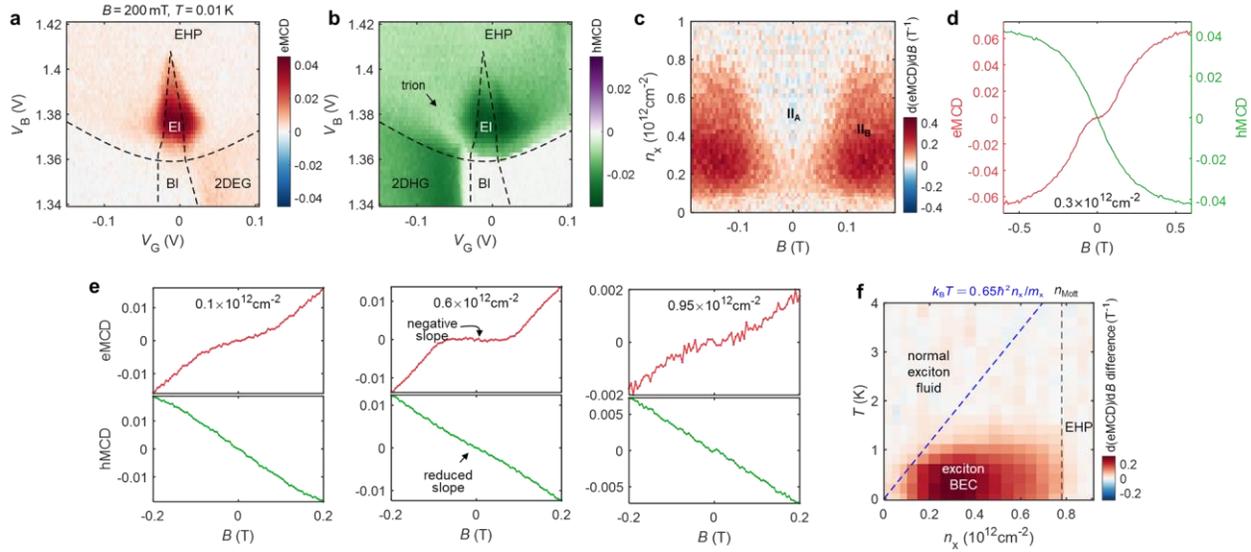

**Extended Data Fig. 8 | Two-component condensate in other TMD material combinations (electron-doped WSe$_2$, hole-doped MoSe$_2$; device D3, 60° angle alignment).**

**a-b**, 2D color plot of eMCD and hMCD as functions of $V_G$ and $V_B$ at 200 mT field. Here eMCD and hMCD are the MCD signal from WSe$_2$ X$^-$ and MoSe$_2$ X$^+$ resonances respectively. A large negative electric field (-0.45V/nm) and a negative bias voltage (positive voltage on MoSe$_2$) reverses the initial type-II band alignment, pushing electrons into WSe$_2$ and holes into MoSe$_2$. A significantly larger $V_B$ is required to overcome the bandgap, and contact performance is worse, both leading to a slightly distorted doping phase diagram. Nonetheless, the MCD enhancement in the EI phase is still clearly observed.

**c**, $B$-derivative of eMCD signal along the charge neutrality line. A similar phase II$_A$ region appears as reversed eMCD slope.

**d**, Comparison of eMCD (left axis) and hMCD (right axis) at exciton density $0.3 \times 10^{12}$ cm$^{-2}$. At the phase II$_B$/I boundary, eMCD and hMCD saturate at the same field.

**e**, Representative eMCD and hMCD at pair densities 0.1, 0.6 and $0.95 \times 10^{12}$ cm$^{-2}$.

**f**, Condensate phase diagram in the density-temperature space. The condensate dome has very similar shape compared to Fig. 4, with similar transition temperature and Mott density.